\documentstyle[preprint,eqsecnum,aps]{revtex}
\tightenlines

\begin{document}
\draft
\title{A Vishniac type contribution to the polarisation of the CMBR?
\\}
\author{T.R. Seshadri$^1$ and K. Subramanian$^{2,3*}$ \\ }
\address{1. Mehta Research Institute, Chhatnag Road, Jhusi,
Allahabad 211019, INDIA}
\address{2. Astronomy Centre, University of Sussex, Falmer, 
Brighton BN1 9QH,UK.}
\address{3. Max Plank Institute for Astrophysics, Karl Schwarzschild 
Strasse 1, 85748 Garching, Germany. }

\date{\today}
\maketitle
\begin{abstract}
Radiation which has a quadrupole component of anisotropy, 
can get polarized by Thomson scattering from charged particles.
In the cosmological context, the microwave background photons
develop significant quadrupole anisotropy  
as they free stream away from the the epoch of standard
recombination.  Reionization in the post recombination era can
provide free electrons to Thomson scatter the
incident anisotropic $CMBR$ photons. We compute the 
resulting polarisation
anisotropy on small (arc-minute) angular scales.
We look for significant non-linear contributions,
as in the case of Vishniac effect in temperature anisotropy,
due to the coupling of small-scale electron density fluctuations, 
at the new last scattering surface, and the temperature quadrupole. 
We show that, while, in cold dark matter type models,
this does not lead to very significant
signals ($\sim 0.02 - 0.04 \mu K$),
a larger small angular scale polarization anisotropy, 
($\sim 0.1 - 0.5\mu K$), can result in isocurvature type models.

\end{abstract}

\narrowtext

\section{Introduction \protect\\}
\label{sec:level1}


The cosmic microwave background radiation has a wealth of 
information about the parameters which govern the dynamics 
and the physical processes in the universe. 
Three important aspects of the CMBR in which the information 
about the universe is encoded are its temperature anisotropy, 
spectral distortion and anisotropy in polarization. 
A study of these aspects will allow one to constrain
the evolution of both the background universe and the 
large scale structures.

Polarization of the CMBR arises from the Thomson scattering of radiation
from free electrons. An analysis of polarization properties of the 
CMBR can reveal valuable information 
about the ionization history of the universe. 
Much of the current work on polarization anisotropy is in the context of
the linear perturbation theory. The expectation generally is that
higher order terms are much smaller. However, for the temperature
anisotropy, on very small angular scales, of order of arcminutes, it was 
shown by Vishniac that nonlinear effects can also make significant
contributions\cite{vish}. These arise through mode coupling of the 
electron density perturbations
on small scales with source terms which vary over larger scales. 
The Vishniac effect is especially important in models
where there is significant early re-ionisation 
(\cite{hss}, \cite{efst88}).

Zaldarriaga\cite{zal} studied the effects of such early
reionisation on first order polarisation anisotropy, in a 
semi-analytical fashion. It turns out that the CMBR
can develop a significant quadrupole, by the epoch of 
re-ionisation, due to the free streaming of the monopole
at recombination. The Thomson scattering of this
quadrupole, off the electrons at the re-ionised
epcoh, can lead to additional polarisation signals.
In this paper we wish to follow suit and 
examine whether this quadrupole, coupling to fluctuations
in the electron density, at the new last scattering surface,
can also lead to significant, Vishniac type second order 
efects. And result in a polarization anisotropy of the 
CMBR at small angular scales. 

In the next section, we give the basic equations and their
formal solution. Section III presents an analytical estimate of the
Vishniac type contribution to the small scale polarization anisotropy,
in re-ionised models. Section IV gives numerical values 
for some illustrative models of structure formation. We
summarise our conclusions in Section V.

\section{Basic equations and their formal solution }

The equations governing the evolution of polarization perturbation
$\Delta_P({\bf x},{\bf \gamma}, \tau)$ and temperature 
perturbation $\Delta_T({\bf x}, {\bf \gamma}, \tau)= \Delta T/T$, 
for scalar modes can be
derived from the moments of the Boltzman equation for photons. 
In the conformal Newtonian gauge, they are given by
\cite{mabert}
\begin{equation}
\dot\Delta_P + \gamma_i\partial_i\Delta_P = 
n_e\sigma_Ta(\tau)(-\Delta_P+
\frac{1}{2}[1-P_2(\mu)]\Pi)  
 \label{eveq}
\end{equation}
\begin{equation}
\dot\Delta_T + \gamma_i\partial_i\Delta_T =
\dot{\phi}-\gamma_i\partial_i\psi+
n_e\sigma_Ta(\tau)(-\Delta_T+\Delta_{T0}
+\gamma_iv_i - \frac{1}{2}P_2(\mu)\Pi)  
\label{eveqt}
\end{equation}
Here ${\bf x}$ is the comoving co-ordinate, $\tau$ is conformal
time, $n_e$ the electron density, {\bf v} the fluid velocity,
${\bf \gamma}$ is the direction of photon propogation, 
$\phi$ and $\psi$ the conformal Newtonian potentials, and
a dot represents derivative with respect to conformal time.
We have also defined
\begin{equation}
\Pi({\bf x},\tau)=-\Delta_{T2}({\bf x},\tau)-
\Delta_{P2}({\bf x},\tau)+\Delta_{P0}({\bf x},\tau) .
\label{pidef}
\end{equation}
with $\Delta_{T0}$, $\Delta_{P0}$ the monoplole, and 
$\Delta_{T2}$, $\Delta_{P2}$ the quadrupole temperature
and polarisation perturbations, respectively. 
We define these angular moments by
\begin{equation}
\Delta_{P,T}({\bf x},{\bf \gamma},\tau) = 
\sum_l (2l + 1)P_l(\mu)
\Delta_{Pl,T l}({\bf x},\tau); \quad
\Delta_{Pl,Tl} = \int {d\mu \over 2} P_l(\mu) \Delta_{P,T}(\mu)
\label{angdel}
\end{equation}
[Here, we make the usual assumption that after Fourier
transformation, the evolution equations for 
temperature and polarisation perturbations,
depend on ${\bf \gamma}$, only in the combination 
${\bf \gamma}.{\bf k}$,
where ${\bf k}$ is the wavevector; so 
we define $\mu = {\bf \gamma}.{\bf k}/k$, where 
$k =\vert {\bf k} \vert$ (cf. ref.\cite{mabert})].

In order to treat inhomogenieties in the electron density,
we take $n_e({\bf x}, \tau) = \bar{n}_e(\tau)[1+\delta_e({\bf x}, \tau)]$,
where $\delta_e$ is the fractional perturbation of electron density 
about the space averaged mean. In general we will have
$\delta_e << 1$.
We may note at this point that spatial perturbations in the 
number density of the electrons is precisely the feature that
gives rise to Vishniac effect in second order temperature perturbations.
We will investigate a similar effect 
for polarization perturbations.

It will be convenient to express equation ~\ref{eveq} 
in terms of the fourier modes as follows,
\begin{equation}
\dot\Delta_{P} + ik\mu\Delta_{P}=
{\bar{n_e}}\sigma_Ta\left[\frac{1}{2}(1-P_2(\mu))
\Pi({\bf k},\tau) - \frac{1}{2}(1-P_2(\mu))
S({\bf k},\tau) 
- \Delta_{P}\right ] 
\label{fourier}
\end{equation}
We have retained the same symbols for the fourier transformed quantities
and defined the mode coupling source term $S({\bf k}, \tau)$ by
\begin{equation}
S({\bf k}, \tau)= -\int  { d^3{\bf p} \over (2\pi)^3} 
\delta_e({\bf k} - {\bf p},\tau)\left[\Pi({\bf p},\tau)
-\Delta_P({\bf p}, \tau) \right] 
\label{capS}
\end{equation}
The formal solution of Eq. (\ref{fourier}) is given by
\begin{equation}
\Delta_P({\bf  k},\tau) = \int^{\tau}_0{\frac{1}{2}}(1-P_2(\mu))
\left[ \Pi({\bf k},\tau') - S({\bf k}, \tau') \right]
g(\tau,\tau')e^{ik{\mu}(\tau'-\tau)}d\tau'  
\label{sol}
\end{equation}
where the $g(\tau,\tau')$ called the visibility function is
given by
\begin{equation}
g(\tau,\tau')=\bar n_e(\tau') \sigma_Ta(\tau')e^{-\int^{\tau}_{\tau'}
\bar n_e(\tau'') \sigma_Ta(\tau'')d{\tau''}}   
\label{g}
\end{equation}
In the above equations, the value of the polarization perturbation 
at the epoch $\tau$ is determined by the entire history 
from $\tau'=0$ to $\tau'=\tau$. The visibility function 
$g(\tau,\tau')$ determines the probability that a photon last 
scatterred at the epoch $\tau'$ reaches us at the epoch $\tau$. 
The exact form of the visibility function is determined by 
the ionization history of the universe. Operationally, 
the role of the visibility function is to give different 
weightages for the integrand for different epochs. 

In this paper we consider a model in which the universe underwent 
a phase of standard recombination and got reionized completely 
at a later epoch, $\tau_{*}$. We will only be concentrating 
on the second order polarization perturbations 
arising from the $S({\bf k}, \tau)$ term in Eq. (\ref{sol}).
The $S({\bf k},\tau)$ contribution to the RHS of
Eq. (\ref{sol}) involves a convolution in the fourier space, which
couples the first order temperature (polarisation) perturbations with 
the first order perturbations in the electron density.
A very similar situation exists
in the case of Vishniac effect in second order temperature 
perturbations. So we expect a Vishniac type effect in second 
order polarization perturbations as well.

Further, the coupling of $\Delta_P$,
$\Delta_{P_0}$ and $\Delta_{P_2}$ with $\delta_e$
in $S$, are likely to be much smaller than the coupling of $\delta_e$ and
$\Delta_{T2}$. This is because firstly the temperature perturabtions
generically dominate the polarisation. Also 
the quadrupole temperature anisotropy, $\Delta_{T2}({\bf k},\tau)$,
will grow to a larger value,
between the epochs of recombination and 
reionisation, due to free streaming of the
monopole at recombination (cf. \cite{zal}). 
We therefore retain in $S$ only
the $\Delta_{T2}$ term and neglect the other terms.
We then have
\begin{equation}
S({\bf k},\tau) \approx \int { d^3{\bf p} \over (2\pi)^3} 
\delta_e({\bf k} - {\bf p},\tau) \Delta_{T2}({\bf p},\tau)    
\label{calS}
\end{equation}
We will adopt this approximate form for the mode
coupling term in what follows.

\section{Vishniac type contribution in reionized models: Analytic estimate}

The absence of "Gunn-Peterson" dips in the spectra
of distant quasars indicates that the universe was probably reionised at
some redshift $ z = z_{*} > 5$. (\cite{gp}).
The value of $z_*$ is
not known observationally, while different theoretical
models have different predictions for this redshift.
In the model which we consider,
the universe underwent standard recombination 
at $\tau_r$, and was reionized completely at a later epoch $\tau_{*}$.
In this case as shown in Ref.\cite{zal},
the visibility function has two peaks, one 
around $\tau_r$ and another peak around $\tau_{*}$. 
We wish to consider here Vishniac type, second order contribution 
to $\Delta_P$. This comes dominantly 
from the value of $\tau'$ around the latter peak. 
It is then convenient to separate the integral over $\tau'$
in Eq. (\ref{sol}) in two parts: $ 0 < \tau' < \tau_{*}$ and
$ \tau_{*} < \tau' < \tau_0$, where $\tau_0$ is the present
conformal time. We write $\Delta_P = \Delta_P^a + \Delta_P^b$, with
\begin{equation}
\Delta_P^a({\bf  k},\tau) = \int^{\tau_*}_0{\frac{1}{2}}(1-P_2(\mu))
\left[ \Pi({\bf k},\tau') - S({\bf k}, \tau') \right]
g(\tau,\tau')e^{ik{\mu}(\tau'-\tau)}d\tau'  
\label{sola}
\end{equation}
\begin{equation}
\Delta_P^b({\bf  k},\tau) = \int^{\tau_0}_{\tau_*}{\frac{1}{2}}(1-P_2(\mu))
\left[ \Pi({\bf k},\tau') - S({\bf k}, \tau') \right]
g(\tau,\tau')e^{ik{\mu}(\tau'-\tau)}d\tau'  
\label{solb}
\end{equation}

The first contribution in (\ref{sola}) is simply 
$\Delta_P^{a} \equiv \exp{(-\kappa_{*})}\Delta_P^{NR}$,
Here $\Delta_P^{NR}$ is the polarisation that would be measured
if there was no reionisation and $\kappa_*$ is the optical
depth to Thomson scattering between now and recombination.
This contribution is reduced by the the fact that only 
a fraction $\exp{(-\kappa_{*})}$ of the photons that arrive at
the observer come directly from recombination, without further 
scattering. (Also the second order contributions
from the $S$ term are much smaller than the first order term
because the electron density fluctuations at recombination
$\delta_e(\tau_r) \sim 10^{-4} - 10^{-3} << 1$, for the relevant
$k$ values).

In order to calculate the second contribution, one has to 
determine the form of the visibility function after 
the standard recombination epoch, that is 
$g(\tau_0,\tau')$ for $\tau' > \tau_r$. 
Using the exact form for $g(\tau_0,\tau')$, to solve for the $\Delta_P$ is not
analytically tractable. So we resort to an approximation for
 $g(\tau_0,\tau')$ in this work, which, while preserving its main features,
also allows analytical results to be derived.
We will return to a full numerical treatment of the problem elsewhere.
In particular, we choose the form of the visibility function after
standard recombination, to be a truncated exponential, given by
\begin{equation}
g(\tau_0,\tau')= N\frac{1}{\sigma}e^{-\frac{(\tau'-\tau_{*})}{\sigma}} 
\theta(\tau'- \tau_{*})
\label{visapp}
\end{equation}
Here  the Heavyside $\theta(x)$ function, is zero for $x < 0$ and $1$
for $x > 0$. It takes account of the fact that before reionisation,
$n_e =0$. Further, $N$ is a normalisation consant and 
$\sigma$ gives the spread of the 
exponential. By appropriately chosing $\sigma$, 
we can set the width of the reionised last scattering surface.
Also note that $g(\tau_0,\tau')$ has the interpretation of probability; so
its integral over $\tau'$ from $\tau'=0$ to $\tau'=\tau_0$ should be normalized to unity. 
This determines the normalisation factor $N$.
For a sufficiently early epoch of reionisation, we 
generally have $(\tau_0 -\tau_{*})/\sigma_2 >> 1$.
In this case, the condition that the integral  of 
$g(\tau_0,\tau')$ over $\tau'$ should be unity implies 
$N + e^{-\kappa_*} =1$,
or $N = 1 - e^{-\kappa_*}$. So $N$ measures the probability
of at least one scattering between $\tau_0$ and $\tau_*$,
due to the reionisation.
Another feature to note is that in our approximation,
$\tau_0$ does not appear at all. This is because, for
the models we will consider, the major contribution
to the scattering optical depth comes from epochs much before 
$\tau_0$.

In the equation (\ref{solb}) for $\Delta_P^b$, 
the first order contribution to the polarisation due to a reionised
universe has already been considered in detail by Zaldarriaga \cite{zal}.
So, here, we concentrate on purely the second order Vishniac type
effect, due to $S$, which we call $\Delta_P^V$. This can be written as
\begin{equation}
\Delta_P^V=-\frac{1}{2}[1-P_2(\mu)]\int_{\tau_*}^{\tau_0}
\exp^{ik{\mu}(\tau'-\tau)}g(\tau_0,\tau')S({\bf k},\tau')d\tau'
\label{d2p2}
\end{equation}
where retaining only the $\Delta_{T2}$ term the mode coupling 
term is given by Eq. (\ref{calS}).
In evaluating the $\tau'$ integral in Eq. (\ref{d2p2}) we assume
the visibility function to be given by the tuncated exponential
form of Eq. (\ref{visapp}).

Let us look at the mode coupling term  $S({\bf k},\tau)$,
given by Eq. (\ref{calS}), in a little more detail. 
This term involves a coupling of 
the quadrupole temperature perturbation at $\tau > \tau_*$, and the
electron density perturbation at the same epoch. Note that
the temperature quadrupole at late times, 
can have a significant contribution due to
the free-streaming of the monopole. For example, in a flat universe,
at large enough wavelengths, the first order quadrupole temperature 
perturbation, is related to the temperature perturbations 
at recombination by \cite{husugsii},
\begin{equation}
\Delta_{T2}(p,\tau_e)=
[\Delta_{T{_0}}(p,\tau_r)+\psi(p,\tau_r)]j_2[p(\tau -\tau_r)]           \label{DeltaTr}
\end{equation}
where $j_2$ is the second order spherical Bessel function,
and $p = \vert {\bf p} \vert$. (Here we have assumed that
$p$ is small enough that the doppler velocity term makes
little contribution to the fre-streamed qudrapole cf.\cite{husugsii}). 
The mode coupling term can then be written as
\begin{equation}
S({\bf k},\tau)=
\int { d^3{\bf p} \over (2\pi)^3} \delta_e({\bf k} - {\bf p},\tau )
[\Delta_{T{_0}}(p, \tau_r)+\psi(p, \tau_r)]j_2[p(\tau -\tau_r)]. 
 \label{caljS}
\end{equation}
The spherical Bessel function, $j_2(\xi)$, is well approximated 
by a gaussian peaked at $\xi\sim3.345$ and with a spread of $\sim 1.4$.
So $j_2[p(\tau -\tau_r)]$ is peaked at wavenumbers around
$p = p_0 \sim 3.345/(\tau -\tau_r)$.  
Note that for the small scale polarisation anisotropy
which we wish to calculate, $k = \vert {\bf k} \vert >> p_0$;
in general we will have $k \sim (10Mpc)^{-1} $ to $(1 Mpc)^{-1}$, 
where as $p_0 \sim (300Mpc)^{-1}$ to $(1000Mpc)^{-1}$, for $\tau > \tau_*$ 
(cf. Ref.\cite{zal} and see below). So in the mode coupling integral,
for a fixed $k >> p_0$, the electron density perturbation 
$\delta_e(({\bf k} -{\bf p}), \tau )$ varies negligibly 
with ${\bf p}$, in the range of
${\bf p}$ for which the the bessel funtion makes significant
contribution. So we can evaluate $\delta_e$ at $p = p_0$ and
pull it out of the $p$ integral. Also since $k >> p_0$, one
can approximate $({\bf k} - {\bf p}_0) \sim {\bf k}$. The
mode coupling ingral, for large $k >> p_0$ then
simplifies to the uncoupled form,
\begin{equation}
S({\bf k},\tau )=\delta_e({\bf k},\tau )
\int { d^3{\bf p} \over (2\pi)^3} \Delta_{T2}({\bf p},\tau )
\equiv \delta_e({\bf k},\tau ) Q_2(\tau)
\label{calSS}
\end{equation}
where we have used Eq. (\ref{DeltaTr}) to rewrite the resulting 
expressions in terms of $\Delta_{T_2}$ again, and defined for
later convenience,
\begin{equation}
Q_2(\tau)= \int { d^3{\bf p} \over (2\pi)^3}
\Delta_{T2}({\bf p},\tau)
\label{q2}
\end{equation}

We now evaluate $\Delta_P^V$, using Eq. (\ref{d2p2}),
and Eq. (\ref{visapp}). Let us assume that the
re-ionisation epoch is late enough, that the
electron density perturbation trace the perturbations
in total matter density. Then, in a flat universe, 
the electron density perturbations grow as $a(\tau) \propto \tau^2 $.
For an open model or one involving a cosmological constant $\Lambda$,
the growth law is given as
\begin{equation}
\delta_e({\bf k}, \tau)
= \left({\tau \over \tau_0}\right)^2 \delta_e({\bf k},\tau_0)
{f(\Omega(z(\tau))) \over 
f(\Omega_0) },
\label{elefl}
\end{equation}
Here, $\tau_0$ is present conformal time and 
the functions $f$  takes into account the reduced growth
in the open or flat $\Lambda$ models at late times, compared to a flat, dark matter
dominated model. In a flat universe $\Omega(z)=\Omega_0=1$, 
$f(1) =1$ and we recover the $\delta_e \propto \tau^2$ growth law.
The expressions for $f$, and $\Omega(z)$ for
$\Lambda$ dominated and the open models 
can be found in ref. \cite{lidlyvw} and  \cite{liddopen},
respectively (in these papers, our $f(\Omega)$ is called $g(\Omega)$).
For $z >> 1$, or $\tau/\tau_0 << 1$ and $f(\Omega(z)) \to 1$.
So the approximation $\delta_e(k,\tau) = (\tau/\tau_0)^2
\delta_e(k,\tau_0) f^{-1}(\Omega_0)$, works very well for
these other models as well, at sufficiently early times. 

Further, as displayed explicitely in ref.\cite{husugsii}, the
power in the CMB monopole, 
per unit logrithmic inteval of $p$ space, 
$p^3[\Delta_{T{_0}}(p,\tau_r)+\psi(p,\tau_r)]$ is
roughly constant on scales $ p < (100Mpc)^{-1}$. (This reflects
the fact that perturbations on scales larger than
the Hubble radius at recombination have not evolved, and
are laid out with constant power, for a scale invariant
initial power spectrum). Recall that the
presence of the $j_2$ term, in the integral over ${\bf p}$,
picks out dominantly contributions from $p \sim p_0 < (300Mpc)^{-1}$,
at any time $\tau > \tau_*$. Now in any realisation, one does
expect some variation of the monopole, as $p$ is varied. 
Nevertheless, because of the constancy of monopole power with $p$,
for $p \sim p_0$, 
one expects the integral term $Q_2(\tau)$
in Eq. (\ref{calSS}), to vary much slower with $\tau$, than
the electron density perturbation $\delta_e({\bf k}, \tau)$,
or the visibility function.
This will especially be so, if the visibility function 
is sufficiently peaked around the reionisation redshift
(for $\sigma$ small enough).  
So when evaluating $\Delta_P^V$, we will assume that 
$Q_2(\tau)$ can be evaluated at some effective $\tau_e \sim \tau_*$,
and pulled out of the integral over conformal time $\tau'$.

The remaining integral, which can be done analytically, 
then gives
\begin{equation}
\Delta^{V}_P=- \frac{1}{2}[1-P_2(\mu)]
\exp^{-ik{\mu}(\tau-\tau_*)}
{ N F(k,\mu) \over (1 - ik\mu\sigma) }
\delta_e({\bf k},\tau_*) Q_2(\tau_e) ,
\label{dvpa}
\end{equation}
where we have taken $(\tau_0 -\tau_*)/\sigma >> 1$, as before,
and defined the factor $F(k,\mu)$
\begin{equation}
F(k,\mu) = 1 + {2\sigma \over \tau_*} {1 \over (1 - ik\mu\sigma)}
+{2\sigma^2 \over \tau_*^2} {1 \over (1 - ik\mu\sigma)^2} .
\label{Fdef}
\end{equation} 
(We have also assumed $\tau_*/\tau_0$ is small enough that
$\Omega(\tau_*) \approx 1$).

We are now in a position to compute the Vishniac type contribution
to the polarisation power spectrum. We define this simply by
analogy to the temperature power spectrum (cf.\cite{efs}).
Recall that, for the temperature perturbations, one first expands 
in spherical harmonics, with
\begin{equation}
{\Delta T \over T}= 
\Delta_T({\bf x}, {\bf \gamma}, \tau_0) = \sum_{l m} a_{lm}
Y_{lm}({\bf \gamma}).
\label{Texp}
\end{equation}
Then the mean square temperature perturbation is
\begin{equation}
<({\Delta T \over T})^2> = \sum_l {(2l +1) \over 4\pi} C_{Tl} \equiv 
\int Q_T(k) {dk \over k}
\label{Tpow}
\end{equation}
where
\begin{equation}
C_{Tl} = < \vert a_{lm} \vert^2> 
= 4\pi \int {k^2 dk \over 2\pi^2} <\vert \Delta_{Tl}(k,\tau_0)\vert^2 >
\label{cldef}
\end{equation}
and 
\begin{equation}
Q_T(k) =  {k^3 \over 2\pi^2} {1\over 2}
 \int^{+1}_{-1}<{\mid}\Delta_T{\mid}^2 >d\mu .
\label{QTdef}
\end{equation}
(Note that, as before, we have taken the normalisation
volume, over which periodic boundary conditions are assumed
to be $V=1$). $Q_T(k)$ gives the power in temperature
perturbations in any logrithmic inteval of $k$.
 
We define therefore, a corresponding Vishniac type
contribution to the polarisation
power spectrum $Q_P(k)$, given by (cf. \cite{efst88}),
\begin{equation}
Q_P^V(k) = {k^3 \over 2\pi^2} {1\over 2}
 \int^{+1}_{-1}<{\mid}\Delta_P^V{\mid}^2 >d\mu .
\label{QPdef}
\end{equation}
We calculate $Q_P^V$ below. 
For this, first take the ensemble average of
$\vert \Delta_P^V\vert^2$. We get
\begin{equation}
<{\mid}\Delta_P^V{\mid}^2 >
=\frac{1}{4}
[1-P_2(\mu)]^2 { N^2 \vert F \vert^2 \over [ 1 + k^2\mu^2\sigma^2]}
P_e(k,\tau_*)\left[ {1 \over 5}
 \left({\Delta T \over T} \right )^2_Q(\tau_e) \right]
\label{polmod}
\end{equation}
In the above we have assumed that the $\delta_e$ and $\Delta_{T2}$
are uncorrelated with each other,
defined the power spectrum of electron density
fluctuation as, $P_e(k,\tau_*) \equiv
 <{\mid}\delta_e({\bf k},\tau_*){\mid}^2>$ and also defined
\begin{equation}
\left({\Delta T \over T} \right )^2_Q(\tau_e) \equiv 
{5C_{T2}(\tau_e) \over 4 \pi} =
5 \int { d^3{\bf p} \over (2\pi)^3}
<{\mid} \Delta_{T2}({\bf p},\tau_e){\mid}^2>.
\label{quaddef}
\end{equation}
Here, $(\Delta T/T)_Q(\tau_e)$ is the quadrupole temperature 
anisotropy as seen by an observer
at the conformal time $\tau_e$.

Now turn to the integral of $ <{\mid}\Delta_P^V{\mid}^2 > $
 over $\mu$. The factor, $1-P_2$ can be expressed as a sum 
of even powers in $\mu$.
\begin{equation}
[1-P_2]^2=\frac{9}{4}(1+{\mu}^4-2{\mu}^2)
\end{equation}
We are generally intersted in the behaviour of the power spectrum 
for large values of $k$, or small angular scales. 
In this case the dominant contribution to the integral over $\mu$, 
in determining $Q_P^V$, will come from the vicinity of $\mu=0$.
It then suffices to retain only the first term in the above
expression for $(1 -P_2)^2$. 
The integral over $\mu$ 
can be done analytically to give the
remarkably simple expression
\begin{equation}
Q_P^V(k,\tau_0) = {9 \pi N^2\over 160} G({\sigma\over\tau_*}) 
\left[{ \Delta_e^2(k,\tau_*)\over k\sigma}\right ] 
\left({\Delta T \over T} \right )^2_Q(\tau_e).
\label{polpowf}
\end{equation}
In doing the integral we have assumed that $k\sigma >> 1$ is
large enough that $\tan^{-1}(k\sigma) \approx \pi/2$,
and defined the factor $G(y) = 1 + 2y + 3y^2 + 5y^3/2 + 35y^4/32$,
where $y=\sigma/\tau_*$. Further,   
\begin{equation}
\Delta_e^2(k,\tau_*) = { k^3 P_e(k,\tau_*) \over 2\pi^2}
= \Delta_e^2(k,\tau_0)\left({\tau_* \over\tau_0}\right)^4
\left({f(\Omega(z_*) \over f(\Omega_0)}\right)^2 
\label{logele}
\end{equation}
is the power per until logrithmic
interval in $k$ space, of the electron density perturbations,
at the epoch $\tau=\tau_*$.

We see that the contribution to polarisation anisotropy,
due to the second order Vishniac type effect, for
re-ionised models, is basically proportional to
the product of the temperature quadrupole, and the
power in electron density perturbations, at last scattering. 
For small angular scales, or large $k$, $Q_P^V$ is 
suppressed because of the finite thickness of
the last scattering surface ($\sigma$), by a factor $k\sigma$.
We note that this suppression is much milder than estimated
in Ref \cite{efst88}, essentially because in that paper,
the first order temperature
quadrupole contribution due to free-streaming of
the monopole at recombination, was not included.
The power spectrum of electron density perturbations
ofcourse depends on the model for structure formation.
Also the parameters $\sigma,\tau_*,N$ depend on the
re-ionisation history. However the power in the temperature quadrupole
at $\tau_e \sim \tau_*$ is likely to be of order the observed
quadrupole; for a large enough $\tau_e$ and if
it arises due to the free-streaming of the monopole at 
recombination. This is again because of the slow variation
of $Q_2(\tau)$ mentioned previously.
We now use Eq. (\ref{polpowf}) to make numerical estimates
of the polarisation due to Vishniac type effect in CDM and
other models of structute formation.

\section{Numerical estimates in different models}

\subsection{ CDM and variants}

Consider first the
case of a standard CDM model (SCDM), with matter density equal to
critical density ($\Omega_0=1$), 
a baryonic contribution $\Omega_b=0.05$, and
a Hubble constant $h=(H_0/100 km s^{-1}Mpc^{-1})=0.5$.
The optical depth to Thomson scattering in a fully ionised,
matter dominated, flat universe is given by 
$\kappa(z)= 0.0418\Omega_B h[(1+z)^{3/2}-1]$. In general, for a
universe with matter density $\Omega_0$, and
assuming $z_* >> 1$, an optical depth
$\kappa_*$ is obtained at a redshift 
\begin{equation}
z_* \approx 97.1\kappa_*^{2/3}\Omega_0^{1/3}
({\Omega_B h \over 0.025})^{-2/3}.
\label{reiored}
\end{equation}
So to have $\kappa_*=1$, in standard CDM, 
we need $z_* \approx 97.1$, and $\kappa_* =0.2$ needs 
$z_* \approx 33.2$. 
Also in a flat matter dominated universe, the conformal time
is related to the redshift by $\tau= \tau_0/(1+z)^{1/2}$,
where $\tau_0 = 2H_0^{-1} = 6000 h^{-1} Mpc$. (Note we
adopt units with c=1). So given $z_*$, this fixes $\tau_*$;
for $\kappa_*=1$, we have $\tau_* \approx 605.8h^{-1}Mpc$,
while for $\kappa_*=0.2$, we have $\tau_* \approx 1025.8h^{-1}Mpc$.

In order to estimate the parameter $\sigma$, in the
model visibility function (\ref{visapp}), we proceed as follows. 
Let $\tau_m=\tau_* +\sigma$. From Eq. (\ref{visapp}), at epochs
after re-ionisation, we have 
$g(\tau_0,\tau_*)/g(\tau_0,\tau_m)= e$. For the exact visibility
function, the same ratio is given by 
$g(\tau_0,\tau_*)/g(\tau_0,\tau_m)= [a^2(\tau_m)/a^2(\tau_*)]
\exp(\kappa_m -\kappa_*)$, where $\kappa_m=\kappa(\tau_m)$.
Equating these two expressions gives an estimate
of $\sigma$. In particular, using $\kappa \propto (1+z)^{3/2}$,
valid for large $z >> 1$ and $a \propto \tau^2 \propto (1+z)^{-1}$,
we have the implicit equation for $\tau_m/\tau_*$,
$1= 4ln(\tau_m/\tau_*) - \kappa_*[1 - (\tau_*/\tau_m)^3]$.
For $\kappa_*=1$, this gives $\sigma \approx 0.54 \tau_*= 327h^{-1}Mpc$,
while for $\kappa_*=0.2$, one gets $\sigma= 0.32\tau_*=328.3h^{-1}Mpc$.

It remains to fix $\Delta_e$ and the temperature quadrupole.
We take the $\Delta_e(k,\tau_0)=\Delta(k)$,
where the matter power spectrum 
\begin{equation}
\Delta^2(k) = {k^3 P(k) \over 2\pi^2} =\left({k \over H_0}\right)^4
\delta_H^2(k)T^2(k)
\label{matpow}
\end{equation}
with the transfer function $T(k)$
in the form given by Bardeen {\it et al.}\cite{BBKS}
\begin{equation}
T(q)= {ln(1+2.34q) \over 2.34q \left[
1+ 3.89q +(16.1q)^2 + (5.64q)^3 + (6.71q)^4\right]^{1/4} }
\label{bbkstr}
\end{equation}
where $q=k/(h\Gamma)$ (cf. ref.\cite{Sugi};\cite{husugi2}).
The parameter $\Gamma$ is referred to as the shape parameter,
is given in ref. \cite{husugi2} (eq. D28 and E12). It is
of order $0.48$ for standard CDM. The four-year COBE normalisation
gives $\delta_H(k=H_0) = 1.94 \times 10^{-5}$
(cf. ref.\cite{bunwhit}), for a scale invariant initial 
power spectrum (with $n=1$). For such initial
conditions, the COBE data also give the 
$(\Delta T/T)_Q(\tau_0) = 18\pm1.6 \mu K$ \cite{cobquad}.
The value of $\tau_* > 1000Mpc$, that we generally
obtain, is likely to be large enough so that
$(\Delta T/T)_Q(\tau_e) \sim (\Delta T /T)_Q(\tau_0)$,
with reasonable accuracy. So we will scale the quadrupole
at $\tau_e$ with the present day observed value.
From the above considerations, and normalising all
quantities to a value of $k=k_c= 1hMpc^{-1}$, we get
\begin{equation}
\sqrt{Q_P^V(k)} = 2.2 \times 10^{-2} \mu K \left({\Delta_e^2(k)/k \over
\Delta^2(k_c)/k_c}\right)^{1/2}
\left({(\Delta T/T)_Q(\tau_e) \over 18 \mu K}\right);\quad 
{\rm for} \quad \kappa_*=1
\label{ncdmk1}
\end{equation}

In case $\kappa_*=0.2$, one has to replace the
numerical value in (\ref{ncdmk1})  by 
$\sqrt{Q_P^V(k_c)} \approx 1.5 \times 10^{-2} \mu K$.
Lower values of $\kappa_* < 1$ may already be
implied by the observed tentative rise in the
CMB anisotropy on degree scales. Note that 
decreasing $\kappa_*$, means a decrease in the
fraction of photons last scattered from the re-ionised
epochs, and so a decrease in $Q_P^V$. But at the
same time since $z_*$ is decreased, the electron
density perturbations at last scattering are larger than the
$\kappa_*=1$ case, which partially compensates by
increasing $Q_P^V$. If the power spectrum can be 
approximated locally as a power law $
\Delta^2(k) \propto k^{3+n}$,
then $Q_P^V(k) \propto k^{2+n}$. Recall that on 
galactic scales with $k\sim k_c$, $n \sim -2$,
while for $k << k_c$, $n \sim -1$ and for $k >> k_c$,
$n \to -3$. So the polarisation anisotropy 
$Q_P^V(k) \propto k^{2+n}$, will increse with $k$
at $k << k_c$, and decrease for $k >> k_c$, attaining
a maximum at $k \sim k_c$. This was one of the reasons for
normalising our estimate to $k =k_c$.

For small angular scales, one can also set up an 
approximate correspondence between the wavenumber $k$
and the angular mutipole number $l$, using $l=kR_*$.
Here $R_*$ translates co-moving distance at the
surface of last scattering, (roughly the epoch when
$\tau=\tau_*$), to angle on the sky, and for a flat model
is given by $R_* = \tau_0 - \tau_*$. In case $\sigma/\tau_0 << 1$,
this approximation will be reasonable, but for
a thick last scattering surface the above correspondence
is less accurate. In the model discussed above, where
$\kappa_*=1$, a wavenumber $k = k_c = 1h Mpc$, corresponds
to $l \approx 5394$, and an angular
scale $\sim 1/l \sim 0.64 $ arc minutes. 

We briefly discuss the predicted small angular
scale polarisation anisotropy $Q_P^V$, for some
variations on the standard CDM model, by using
the same method of calculation as above. For example,
increasing the baryon density to
$\Omega_b=0.1$, leads to a smaller redshift
of last scattering with a given $\kappa_*$ but
also a smaller $\Gamma$.
For $\kappa_*= 1$, one gets a slightly larger value 
$\sqrt{Q_P^V(k_c)} \sim 2.9 \times 10^{-2} \mu K$,
than in SCDM. Also
if the primordial spectrum is tilted to $n=1.2$,
the best fit slope determined by COBE \cite{cobquad},
(keeping all other parameters of SCDM same), 
then also one has a larger value $\sqrt{Q_P^V(k_c)}
\sim 4.0 \times 10^{-2} \mu K$. 

Suppose we adopt a $\Lambda +$ CDM 
type model, as discussed for example in Ref.\cite{ostein}, 
with $\Omega_0=0.35$, $\Omega_{\Lambda}=0.65$, $h=0.7$, $n=1$,
$\Omega_b=0.04$. Then we get $z_* \sim 63$ for $\kappa_*=1$.
Assuming that $\tau_*$ is early enough that $\sigma$
can be estimated as for the flat universe, we have 
\begin{equation}
\sigma = {6000h^{-1} Mpc \over \Omega_0^{1/2}(1 + z_*)^{1/2}}
\left({\tau_m \over \tau_*} -1 \right)
\label{sig}
\end{equation}
So for the $\Lambda +$ CDM model, we get $\sigma \sim 687.5h^{-1}Mpc$.
For this model $f^{-1}(\Omega_0) \sim 1.24$, and 
adopting a normalisation of
the power spectrum as in \cite{bunwhit,lidlyvw}
we then get $\Delta_e(k_c,z_*) \sim 3.56/(1+z_*)$, which leads to 
$\sqrt{Q_P^V} \sim 1.9 \times 10^{-2} \mu K 
((\Delta T/T)_Q(\tau_e)/18\mu K)$.
 Note that in this model the integrated Sachs-Wolfe effect,
which makes a contribution to the present day quadrupole,
will make little contribution at redshifts $\sim z^*$.
However, using Eq. (10) of ref.\cite{huwhit}, we estimate
that this will cause only a $10\%$ reduction in the above value. 
For an open CDM model (OCDM), with $\Lambda =0$, but all other
parameters as for the above $\Lambda+$ CDM model, $z_*$ is the same 
and $\Delta_e(k_c)$ is of the same order,
using the power spectrum as determined by \cite{liddopen}.
However the quadrupole at last scattering is 
likely to be smaller; because the integrated Sachs Wolfe effect
contributes a larger part of present day quadrupole.
So the predicted $\sqrt{Q_P^V(k_c)}$ is likely to be smaller
than the for the above models.

\subsection{ Isocurvature type models}

Finally, consider as an alternative to the standard models,
the isocurvature model recently discussed by Peebles \cite{peeb97};
where density perturbations are provided by CDM that is
the remnant of a massive scalar field frozen from
quantum fluctuations during inflation. The novel
feature of such a picture, as pointed out in
\cite{peeb97}, is that the primeval CDM
mass distribution is proportional to the square
of a random Gaussian process; so prominent
upward fluctuations are much larger (by factor $F \sim 3$), than
for a Gaussian process with the same RMS. The merits of
such a picture has been discussed by Peebles \cite{peeb97}. 
We consider two representative models.
Model 1 discussed in \cite{peeb97} adopts $\Omega_0=0.3$, $\Lambda=0.7$, 
$\Omega_b=0.05$, $h=0.7$ and, a matter power spectrum, 
which, on the relevant small scales, can be approximated as 
$\Delta_e^2(k)=(k/k_0)^{3+m}$, with $m=-1.8$, $k_0=0.1hMpc^{-1}$. 
And for model 2, $\Omega_0=0.1$, $\Lambda=0.9$, 
$\Omega_b=0.05$, $h=0.7$ and, $m=-1.4$.

Note that in these models, due to early structure formation,
re-ionisation is expected to occur at large redshifts.
The optical depth to electron scattering, measured from
the present epoch could then rise to values larger than
unity. However the possible ionisation history in these models
is largely unexplored. In order to get a preliminary estimate of
the anisotropies in polarisation that could be generated, we
simply use Eq. (\ref{polpowf}) (implicitely making the
simplifying assumptions which went into its derivation). 
So the universe after standard recombination
is assumed to be largely neutral, and then re-ionised after 
an epoch $\tau=\tau_*$. Again at $\tau \sim \tau_*$, a 
quadrupole would arise from the free-streaming of the 
large scale entropy perturbation at recombination 
(the isocurvature effect cf. \cite{sugsil}).
Further, in Eq. (\ref{visapp})
we assume $\tau_*$ to be the epoch with $\kappa_* =1$, but
take $N \sim 1$ (to reflect the fact that little
of the small angular scale anisotropy is due to
conventional last scattering at around the re-combination epoch).
We hope to return to a better treatment of these models
in future work. 

For model 1 of Peebles\cite{peeb97}, one then gets
$z_* \sim 52$, and from Eq. (\ref{sig}), $\sigma = 821h^{-1}Mpc$. 
Also for $z_* >> 1$, we have $f(\Omega(z_*)) \to 1$, while  
for $\Omega_0=0.3$, $\Lambda=0.7$, $f^{-1}(\Omega_0) \sim 1.3$.
Putting in all the numerical values,  
we then get 
\begin{equation}
\sqrt{Q_P^V(k)} \approx 2.7 F \times 10^{-2} \mu K
{(\Delta T/ T)_Q(\tau_e) \over 10\mu K} 
\left({k\over 1h Mpc^{-1}} \right)^{0.1}
\label{isoqp1}
\end{equation}
Here we have scaled $(\Delta T/T)_Q(\tau_e)$ by a
smaller value of $10 \mu K$, since these isocurvature models
predict a somewhat smaller quadrupole than  SCDM models 
(cf. Fig 1. of \cite{peeb97}). We have also incorporated
a factor $F$, to remind ourselves that the non-gaussian
statistics of the density field, may lead to $F$ times larger 
prominent upward fluctuations.  If we wish to 
convert $k$ to $l$ in this model, 
we again use $R_* = \tau_0 - \tau_*$, with 
$\tau_0 \sim 2.17/(\Omega_0^{1/2} H_0)
 \sim 11,862.2 h^{-1} Mpc$ (cf. Eq. (20) of ref.\cite{huwhdamp}).
Also $\tau_* \sim 1520.6 h^{-1}$ and so $k=1h Mpc$
corresponds to $l \sim 10,342$ or an angular scale
$\sim 0.33$ arc minutes.

A similar analysis for model 2, gives $z_* \sim 36$, 
$\sigma \sim 1710h^{-1}Mpc$, and a larger polarisation signal, with
\begin{equation}
\sqrt{Q_P^V(k)} \approx 5.7 F \times 10^{-2} \mu K
{(\Delta T/ T)_Q(\tau_e) \over 10\mu K} 
\left({k\over 1h Mpc^{-1}} \right)^{0.3}
\label{isoqp2}
\end{equation}

Note that Peebles adopts a cosmological constant for convenience
of analysis. If we were to consider open versions
of these models, $z_*$ and $\sigma$ are nearly the same
(since $z_* >> 1$, but the power in electron density 
perturbation at last scattering is much larger, because of
a much larger $f^{-1}(\Omega_0)$. The numerical value, in
Eq. (\ref{isoqp1}) and (\ref{isoqp2}), at
$k=1h Mpc^{-1}$ gets increased respectively to $\sqrt{Q_P^V(k_c)} \sim 
4.6 F \times 10^{-2} \mu K$ (model 1) and 
$\sqrt{Q_P^V(k_c)} \sim 0.17 F \mu K$ (model 2). The
$k$ dependence remains the same. In the open models,
the density on scales of $k=k_c$ are already going non-linear
at $z_*$ and so the above numbers provide only a crude estimate.
We see in general that these isocurvature models predict
much larger polarisation
signals compared to the SCDM type models.
First, the RMS vaue is larger. Further,
because of the non-gaussian statistics of the
density field, one expects
prominent upward fluctuations $F \sim 3$ times larger than
the RMS value (cf. \cite{peeb97}).

We note in passing that the
older versions of the baryonic isocurvature models, 
say with $\Omega_0 \sim 0.2$, $\Omega_b= 0.05$, $h=0.8$,
$m \sim -0.5$ cf. \cite{oldpib}, leads to even
larger signals with $\sqrt{Q_P^V} \sim 0.3 \mu K
((\Delta T/T)_Q(\tau_e)/10\mu K)(k/k_c)^{3/4}$. However these
models may already be ruled out by the fact that they result in
spectral distortions, larger than the limit
implied by the COBE observations \cite{hubunsug}.

\section{Conclusions}

In this paper we have explored the possibility of
a Vishniac type contribution to the polarisation
anisotropy at small angular scales. It is well known 
that non-linear effects can make
significant contribution to temperature anisotropy on
small angular scales, through the Vishniac effect, especially
in re-ionised models. This arises due to the mode coupling of
large angular scale, first-order velocity perturbations, with small
angular scale electron density perturbations. 
We have considered here whether 
a similar effect contributes to the polarization anisotropy, 
by studying the coupling of large angular scale, 
first-order temperature anisotropy (quadrupole) with small
angular scale electron density perturbations, in re-ionized models.

We find that in cold dark matter models and its
variants, the Vishniac type effect leads to a fairly small
polarisation anisotropy, with 
$\sqrt{Q_P^V}\sim 0.02 - 0.04 \mu K$, on scales with 
$k \sim 1h Mpc$ (or angular scales of arc minute or smaller). 
However in isocurvature type models the Vishniac type contribution 
can result in much larger signals. For the models of 
Ref.\cite{peeb97}, the anisotropy on small anugular scales
is non-gaussian, with prominent upward fluctuations of order
$ 0.1 - 0.5 \mu K$, assuming $F \sim 3$. This reflects basically
the fact that, the isocurvature type models have much more power on
small scales and so produce much larger electron density
fluctuations. We note in passing, 
that the suppression factor, due to the
finite thickness of the last scattering surface, on
the small scale polarization anisotropy, is much milder than
that obtained in ref.\cite{efst88}. This is because,
as mentioned earlier, the first order temperature quadrupole 
contribution, arising due to the free-streaming of the monopole 
at recombination, was not included in their analysis.

It is clear that the polarisation signals on arc minute
scales predicted by CDM type models will be difficult to detect,
but those predicted by isocurvature type models will be much easier.
If small scale polarisation anisotropy is eventually detected,
it will open up the novel prospect of studying directly, 
both the quadrupole anisotropy and small scale 
electron denity fluctuations, at high redshifts. As pointed
out in ref\cite{loeb}, in a different context, one can then also
reduce the cosmic variance of the quadrupole significantly.
In this paper we have made several approximations to 
analytically estimate the polarization anisotropy.
We plan to return to a better numerical analysis in the near future.

\acknowledgments

TRS and KS thank John Barrow and the Astronomy Centre, University
of Sussex for hospitality during the course of this work. 
KS was supported by a PPARC Visiting Fellowship at the Astronomy
centre, University of Sussex. KS also thanks 
Simon White and the Max-Plank Institute for Astrophysics,
Garching, for hospitality when the paper was being completed, and
Scott Dodelson for a useful discussion.

\bigskip

* On leave from National Centre for Radio Astrophysics, TIFR Pune,
Poona University Campus, Ganeshkhind, Pune 411 007, India.

\end{document}